\documentclass[]{article} 

\title{Noether Theorem and the quantum mechanical operators}

\author{Wai Bong Yeung \\
 phwyeung @ccvax.sinica.edu.tw \\
 Institute of Physics, Academia Sinica, Taipei, Taiwan ,ROC}

\begin{document} 
\maketitle 

\begin{abstract}
We show that the quantum mechanical momentum and angular 
momentum operators are fixed by the Noether theorem for the
classical Hamiltonian field theory we proposed.
\end{abstract} 

Recently, we have proposed a classical Hamiltonian field theory [1],which
in the limit of very large Planck frequency,mimics many aspects of a
quantum
mechanical system. In particular, the Schrodinger Equation will follow
from
the Hamilton's Field Equation, and the Hamiltonian of the classical
field theory will become the energy expectation value of the
corresponding
quantum mechanical system.
    
The Hamiltonian density for the classical field theory we proposed
	  contains 
	  (1+2n) pairs of
	  canonical conjugate variables ($p,q$) ( $P_j , Q _j$), ($\pi_j
	  ,\eta_j$  ), j=1,...,n.All of these canonical variables are
	  functions of  $x=(x_1 ,.. ,x_n )$ and t.And it reads as
	  
	  \begin{equation}
	   H=(1/2h)( V(x))(p ^2 + q ^2 )-(1/2h)(mc^2)(P_j^2   +Q_j^2  +\pi_j^2
	    + \eta_j^2 )
	   -(c/2)p\partial  _j (Q _j+\eta_j )-(c/2)(P _j +\pi_j    )\partial
	   _j q 
	   \end{equation}
     Independent variations of the field variables generate the Hamiltonian 
     Field Equation. If we are interested in the case in which the field 
     variables oscillate with frequencies far smaller than the Planck
     frequency
      h/mc, then the variables {$P_j,Q_j$} and {$\pi_j,\eta_j$} will be
      related to {$p,$q} through [1]

      \begin{eqnarray}
      Q_i &= & h/2mc \partial _i p        \nonumber\\
      P_i &= & -h/2mc \partial _i q         \nonumber\\
      \eta_i & =& h/2mc \partial _i  p         \nonumber\\  
      \pi_i&=& -h/2mc \partial _i  q         ,  j=1,..n
      \end{eqnarray}

     In this paper, we will explore the translational and rotational
     symmetries of this 
     classical Hamiltonian field theory. It is well known in the literature
     that these
     continuous symmetries will lead to conserved physical quantities ; a
     result called
     the Noether theorem [2] .And the aim of this paper is to understand how
     these 
     conserved physical quantities coming from a classical field theory are
     related to the
     measurable quantities of the corresponding quantum mechanical system.                                                                                                                                     

          Let us first consider the $ V(x)= 0 $ case.When there is no external
	  potential present, 
	  there will be both translational and rotational symmetries for the
	  classical field system.
	  And by Noether theorem, there exist some corresponding conserved
	  quantities. 
	  For the translational invariance, the resulting  conserved
	  quantities are the components of the
	  second rank tress-energy tensor $T^{\mu} _{\nu}$,given by Noether as

\begin{equation}
T^{\mu} _{\nu} =\Sigma (\partial L/\partial(\partial_{\mu}u)
)(\partial_{\nu}u)-L\delta^{\mu} _{\nu}
\end{equation}
where L is the underlying Lagrangian density for the classical field theory. u
stands
collectively for all the field variables.Noether theorem requires the
conservation law

\begin{equation}
\partial_{\mu}T^{\mu} _{\nu} = 0
\end{equation}

     We are particularly interested in the vector $ m_{j }$ defined as

     \begin{equation}
     m_j=\int d^nx T_{0j}
     \end{equation}

     These $m_{j}$,other than a multiplicative constant that we shall fix
     later,are always taken
     as the mementum components carried by the classical fields because they
     are generated  by
     the translational symmetries. A close look at $ T^{0} _{j}$ will show
     that they are independent
     of the detailed structures of the Lagrangian density L and has the simple
     form of

\begin{equation}
T^{0} _{j}= \Sigma (\partial L/\partial (\partial_t{u}))(\partial_ju)
              =p\partial_jq +P_i\partial_jQ_i +\pi_i\partial _j\eta_i
\end{equation}
   
Using the result given in Eq(2),$ T^{0} _{j} $ can be written in terms of $p$
and $q$ as

\begin{equation}
T^{0} _{j}= p\partial_jq -2(h/2mc)^2\partial_ip\partial_j\partial_iq
\end{equation}

For a very large Planck frequency , the second term of Eq(7) drops out, and
hence

\begin{equation}
m_j=\int d^nx(-p\partial_jq)
\end{equation}

     If we define the corresponding quantum mechanical wave function  by 
     $\psi(x,t)$=($q(x,t)+ip(x,t)$)/$\sqrt{2}$[1]. It can be seen
     immediately that

     \begin{equation}
     m_j=\int d^nx\psi*(-i\partial_j)\psi,
     \end{equation}

     after integration by parts.

The physical meaning of the above result is the following: If we use
$\star{p}_{j}$ to denote 
the quantum mechanical operator for the j the component of the momentum,
and if we use the 
above $\psi$ to compute the expectation value of the momentum
components, then $\star{p}_{j}$
must be of the form

\begin{equation}
\star{p}_{j}=-i\beta\partial_j
\end{equation}

     where $\beta$ is a proportional constant that will be shown to be h
     later.This result can
     be regarded as a derivation of the most fundamental quantum mechanical
     prescription

\begin{equation}
\star{p_j}=-ih\partial_j
\end{equation}

For the rotational invariance, we assume that $p$,$q $,$P_j$,$Q_j $,$\pi_j$
and $\eta_j$ all transform
as scalars under the rotation group.The resulting conserved quantities will
then be the components of the third
rank angular momentum tensor$M^{\beta}_{\lambda\mu}$, given by Noether as

\begin{equation}
M^{\beta}_{\lambda\mu} = x_{\mu }T^{\beta} _{\lambda} -x_{\lambda}T^{\beta}
_{\mu}
\end{equation}

The components of this third rank tensor that are related to the angular
momentum components of
the classical fields are $M^{0}_{lk}$.Using the result given in Eq(7), it can
be shown easily
that the integrated components 

\begin{equation}
L_{lk}=\int d^nxM^{0}_{lk}
\end{equation}

can be written as 

\begin{equation}
L _{lk}= \int d^nx\psi*(-ix_l\partial_k+ix_k\partial_l)\psi
\end{equation}

And hence the orbital angular momentum $\star{L}$ in quantum mechanics will have
the familiar
form

\begin{equation}
\star{L}=r X \star{p}
\end{equation}

     In the presence of the potential V(x), we will no longer have
     translational invariance, and 
     so no more conservation law. Instaed we shall have [3]

\begin{equation}
d/dx_{\mu}(T^{\mu}_{\nu}) =-\partial_{\nu}L=\partial_{\nu}H
\end{equation}

     The integrated spatial parts for the above equation read as

\begin{equation}
-\partial_t \int d^n xT_{0j}  +\int d^nx d/dx_l T_{lj} = \int
d^nx\partial_jH
\end{equation}

Throwing away the surface term,and using the results given in Eq(1),Eq(8)
and Eq(9), 
Eq(17) will become

\begin{equation}
-\partial_t\int d^nx\psi*(-i\partial_j)\psi = 1/h\int
     d^nx\psi*(\partial_jV)\psi
\end{equation}

or

\begin{equation}
\partial_t \int d^nx\psi*(-ih\partial_j)\psi =\int d^nx
\psi*(-\partial_jV)\psi
\end{equation}

     This is the Ehrenfest theorem [4] that we always encounter in quantum
     mechanics. And as we have 
     promised before, we have fixed the proportional constant $\beta$ that
     appeared in Eq(10) to be the Planck constant h.

          An equation similar to Eq(18) can also be derived for the angular
	  momentum which relates the
	  rate of change of angular momentum with the external applied torque.

	       So we may conclude our paper by saying that the quantum
	       mechanical operators for the momentum
	       and angular momentum variables will be fixed by the Noether
	       theorem for our classical Hamiltonian
	       field theory.

\end{document}